\documentclass[twocolumn,showpacs,preprintnumbers,amsmath,amssymb,prb,superscriptaddress]{revtex4-1}
\usepackage{mathrsfs}
\usepackage{graphicx}
\usepackage{dcolumn}
\usepackage{bm}
\usepackage{amsmath}
\usepackage{amsfonts}
\usepackage{color}

\begin{document}

\title{Edge State Induced Andreev Oscillation in Quantum Anomalous Hall Insulator-Superconductor Junctions}
\author{Biao Lian}
\affiliation{Department of Physics, McCullough Building, Stanford University, Stanford, California 94305-4045, USA}
\author{Jing Wang}
\affiliation{State Key Laboratory of Surface Physics and Department of Physics, Fudan University, Shanghai 200433, China}
\affiliation{Department of Physics, McCullough Building, Stanford University, Stanford, California 94305-4045, USA}
\affiliation{Stanford Institute for Materials and Energy Sciences, SLAC National Accelerator Laboratory, Menlo Park, California 94025, USA}
\author{Shou-Cheng Zhang}
\affiliation{Department of Physics, McCullough Building, Stanford University, Stanford, California 94305-4045, USA}
\affiliation{Stanford Institute for Materials and Energy Sciences, SLAC National Accelerator Laboratory, Menlo Park, California 94025, USA}

\begin{abstract}
We study the quantum Andreev oscillation induced by interference of the edge chiral Majorana fermions in junctions made of quantum anomalous Hall (QAH) insulators and superconductors (SCs). We show two chiral Majorana fermions on a QAH edge with SC proximity generically have a momentum difference $\Delta k$, which depends on the chemical potentials of both the QAH insulator and the SC. Due to the spatial interference induced by $\Delta k$, the longitudinal conductance of QAH-SC junctions oscillates with respect to the edge lengths and the chemical potentials, which can be probed via charge transport. Furthermore, we show the dynamical SC phase fluctuation will give rise to a geometrical correction to the longitudinal conductance of the junctions.
\end{abstract}

\date{\today}


\maketitle

Quantum anomalous Hall (QAH) state is known as a two-dimensional (2D) topological state which has an integer number $N_h$ of chiral fermions at the edge and exhibits a quantized Hall conductance in the absence of an external magnetic field~\cite{Haldane1988,Liu2008,Qi2008,Hasan2010,Qi2011,Yu2010,Wang2013a,Onoda2003,Wang2013b,Wang2015R,Liu2016}. For non-interacting fermionic systems, $N_h$ is the total Chern number of the occupied electronic bands. The QAH state with $N_h=1$ has been experimentally realized in both Cr-doped~\cite{Chang2013,Kou2014,Checkelsky2014,Bestwick2015} and V-doped~\cite{Chang2015} (Bi,Sb)$_2$Te$_3$ magnetic topological insulator thin films. When the QAH state is proximity-coupled with a normal $s$-wave superconductor (SC), the system becomes a chiral topological SC (TSC) and the edge chiral Majorana fermions arise~\cite{Schnyder2008,Fu2008,Sau2010,Alicea2010,Qi2010,Rontynen2015}. Such systems may exhibit exotic transport phenomena
due to the existence of electrically neutral Majorana edge states~\cite{Fu2009,Akhmerov2009,Chung2011,Liu2011,Strubi2011,Wang2015,Yamakage2014,He2014}.
However, not much effort has been made to understand how exactly the electric current flows from a QAH insulator into an adjacent normal SC (or TSC), both of which are conductive and dissipationless. This is crucial to the study of coupled QAH/SC transport experiments.

In this Letter, we show the conductance of a QAH/SC junction exhibits an Andreev oscillation due to the interference of the chiral Majorana fermions on the QAH edge proximity-coupled to the SC. Such an interference is induced by the momentum difference $\Delta k$ between the two chiral Majorana fermions on the same edge, which can be tuned by the chemical potentials of both the QAH insulator and the SC. As a result, the two-terminal longitudinal conductance of the QAH/SC junction oscillates with respect to the length of the proximity-coupled edge and the chemical potentials of QAH and SC, while the Hall conductance is quantized. Similar Andreev oscillation in the longitudinal conductance occurs for the other junctions of QAH insulator and SC shown in Fig.~\ref{Fig3}, while the Hall conductance always remains quantized. Furthermore, we consider the QAH/TSC/QAH junction, where there is only a single chiral Majorana fermion on each superconducting edge. The dynamical phase fluctuation of SC will have a $1/d_{\text{SC}}^3$ geometric correction to the previously predicted half-quantized longitudinal conductance $e^2/2h$~\cite{Chung2011,Wang2015}, where $d_{\text{SC}}$ is the size of TSC in the junction, $e$ is the electron charge and $h$ is the Plank constant. All the conclusions discussed here also hold for integer quantum Hall (IQH) insulator/SC junctions.

\begin{figure}
\begin{center}
\includegraphics[width=3.3in]{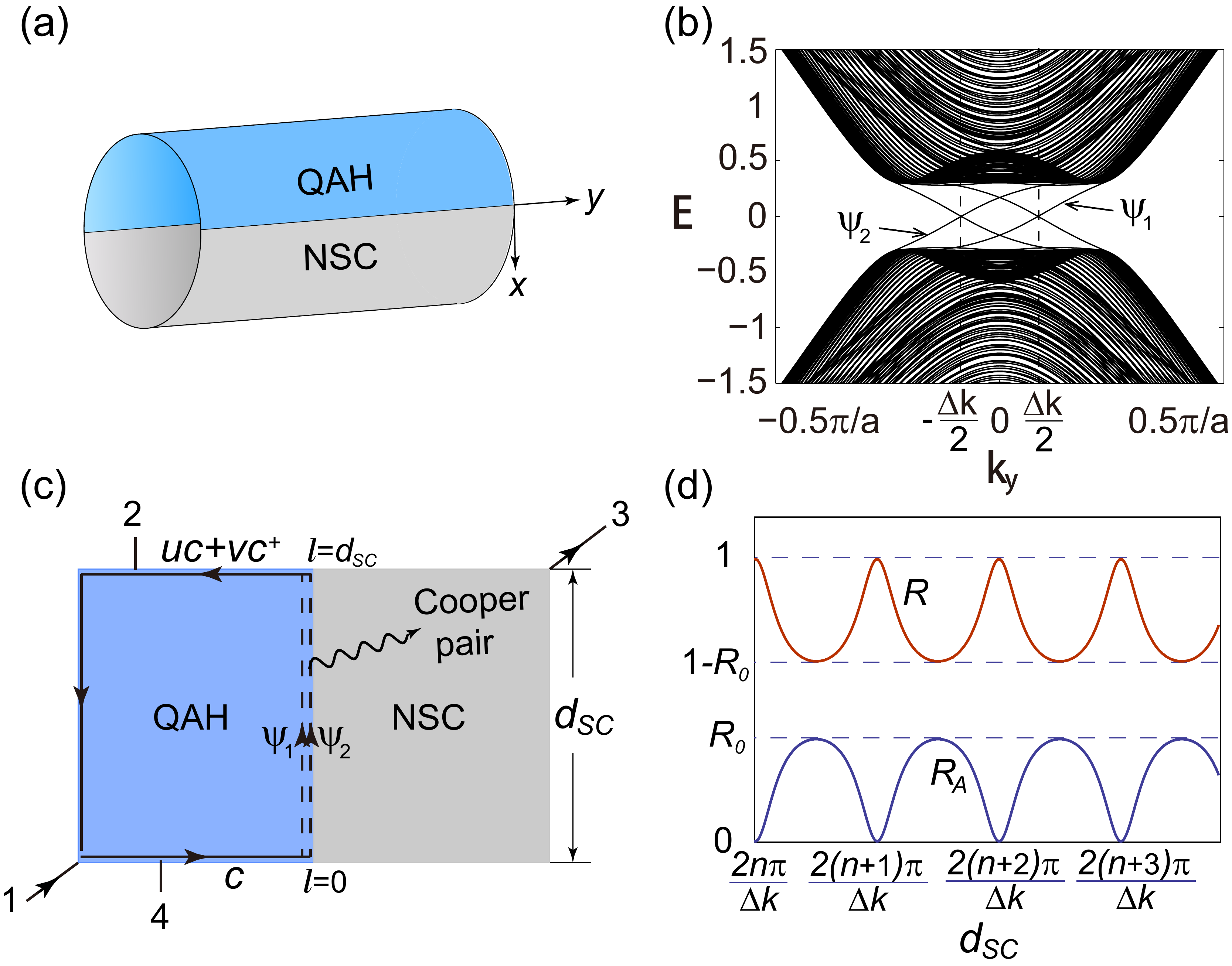}
\end{center}
\caption{(color online). (a) The QAH/NSC junction in cylinder geometry. (b) The BdG spectrum of the junction in (a), where the two chiral Majorana modes have a momentum difference $\Delta k$. (c) Illustration of a QAH/NSC junction with an edge length $d_{\text{SC}}$. (d) The Andreev reflection probability $R_A$ and the normal reflection probability $R$ of the junction with respect to $d_{\text{SC}}$.
}\label{Fig1}
\end{figure}

The basic mechanism of the edge chiral Majorana fermions interference in a QAH/SC junction can be easily understood in the geometry shown in Fig. \ref{Fig1}(a), where a QAH insulator and a normal SC (NSC) are attached into a $y$-direction translational invariant cylinder. Since a QAH with Chern number $N_h$ is topologically equivalent to a chiral TSC with Bogoliubov-de Gennes (BdG) Chern number $N=2N_h$, the $N_h$ chiral fermions on the QAH edge will become $2N_h$ chiral Majorana fermions under the proximity effect of the NSC. For simplicity, we restrict ourselves to QAH with $N_h=1$. In this case, the two chiral Majorana fermions on the same QAH edge are related to each other by the particle-hole symmetry (PHS). In general, the energy dispersions of these two chiral Majorana fermions will not coincide with each other. To show this, we take the two-band lattice Hamiltonian for the QAH:
\begin{equation}\label{HQAH}
\mathcal{H}_{\text{QAH}}=\sum_{\mathbf{k}}c^\dag_\mathbf{k}\left[\bm{\zeta}(\mathbf{k})\cdot\bm{\sigma}-\mu_{h}\right]c_{\mathbf{k}}\ ,
\end{equation}
and the $s$-wave BdG Hamiltonian for the NSC:
\begin{equation}\label{HNSC}
\mathcal{H}_{\text{NSC}}=\sum_{\mathbf{k}}c^\dag_\mathbf{k}\left[\epsilon(\mathbf{k})-\mu_{s}\right]c_{\mathbf{k}}+(\Delta_s c_{\mathbf{k}}^T i\sigma_yc_{-\mathbf{k}}+\mathrm{H.c.})\ .
\end{equation}
Here, the basis $c_{\mathbf{k}}=(c_{\mathbf{k}\uparrow},c_{\mathbf{k}\downarrow})^T$, $\bm{\zeta}(\mathbf{k})=(M-B(\cos k_xa+\cos k_ya), A\sin k_x a, A\sin k_y a )$, $\bm{\sigma}=(\sigma_x,\sigma_y,\sigma_z)$ are the Pauli matrices, $\epsilon(\mathbf{k})=B(2-\cos k_xa-\cos k_ya)$ is the kinetic energy, $\mu_{h}$ and $\mu_s$ are the chemical potentials of the QAH and the NSC, respectively, $a$ is the lattice constant, and $\Delta_s$ is the pairing amplitude. The QAH insulator is realized in the regime $|M|<2|B|$ and $|\mu_h|<2|B|-|M|$. In Fig.~\ref{Fig1}(b), the BdG spectrum of the cylinder is calculated as a function of $k_y$ with parameters $a=0.8$, $B=1.5625$, $M=2.625$, $A=1.25$, $\Delta_s=0.3$, $\mu_h=0.2$ and $\mu_s=0.5$. The distinction between the dispersions of two chiral Majorana fermions $\psi_{1}$ and $\psi_{2}$ on the same edge is clearly seen, where the momentum difference between $\psi_{1}$ and $\psi_{2}$ at zero energy is denoted as $\Delta k$.

Now we consider a QAH/NSC junction as shown in Fig.~\ref{Fig1}(c), where the length of the QAH edge (the right edge) in contact with NSC is $d_{\text{SC}}$. The low energy physics in the QAH is dominated by the gapless edge electrons. When an edge electron denoted by $\bar{c}_k$ in the lower edge enters into the right edge of the QAH, it splits into two chiral Majorana fermions $\psi_1$ and $\psi_2$. Whenever $\psi_1$ and $\psi_2$ have a momentum difference $\Delta k$, a phase difference $\phi=\Delta k d_{\text{SC}}$ will be accumulated between them after propagating along the edge of length $d_{\text{SC}}$. For $\phi\neq2m\pi\ (m\in\mathbb{Z})$, the outgoing state in the upper edge will become a superposition of electron and hole $u\bar{c}_k+v\bar{c}_k^\dag$, where $|u|^2+|v|^2=1$ due to the unitarity. Therefore, an incident electron from the lower QAH edge has a probability $|v|^2$ turning into a hole at the upper QAH edge, which is denoted as the Andreev reflection probability $R_A=|v|^2$. Accordingly, the normal reflection probability is $R=|u|^2=1-R_A$. $R_A$ can be calculated by solving a 2D Shr\"{o}dinger equation numerically~\cite{supplement}. Here we give an approximate expression for $R_A$ via a simplified picture as follows. Due to the PHS, the two edge chiral Majorana modes $\psi_{1,2}$ at zero energy take the generic form
\begin{equation}
\psi_1=\alpha \bar{c}_{\frac{\Delta k}{2}}+\beta \bar{c}_{-\frac{\Delta k}{2}}^\dag\ ,\quad \psi_2=\beta^* \bar{c}_{-\frac{\Delta k}{2}}+\alpha^* \bar{c}_{\frac{\Delta k}{2}}^\dag\ ,
\end{equation}
where $|\alpha|^2+|\beta|^2=1$, while $\bar{c}_{k}$ and $\bar{c}_{k}^\dag$ are the edge electron annihilation and creation operators, respectively. When $\Delta_s=0$, we recover the QAH edge state and get $\alpha=1$, $\beta=0$. For convenience the QAH edge is parameterized as $\ell$, where the origin $\ell=0$ is set at the lower right corner of QAH. The chiral edge mode for an incident electron with momentum $k_I$ is then $\Psi(\ell)=\bar{c}_{k_I}=\bar{c}(\ell)e^{ik_I\ell}$ on the lower edge $\ell<0$, and $\Psi(\ell)=u\bar{c}(\ell)e^{ik_I\ell}+v\bar{c}^\dag(\ell)e^{-ik_I\ell}$ on the upper edge $\ell> d_{\text{SC}}$. The vanishing hole probability at $\ell=0$ requires $\Psi(\ell)=\mathcal{N}[\alpha^*\psi_1(\ell)-\beta\psi_2(\ell)]$ on the right edge $0<\ell< d_{SC}$, where $\mathcal{N}$ is a normalization factor. The continuity condition for $\Psi(\ell)$ at $\ell=d_{\text{SC}}$ of junction is $\Psi(d_{SC}^+)\propto\Psi(d_{SC}^-)$, then the Andreev reflection probability $R_A=|v|^2$ is found to be~\cite{supplement}:
\begin{equation}\label{RA}
R_A(\phi)=\frac{4|\alpha\beta|^2\sin^2(\phi/2)}{(|\alpha^2|-|\beta^2|)^2+8|\alpha\beta|^2\sin^2(\phi/2)}\ ,
\end{equation}
with $\phi=\Delta k d_{\text{SC}}$. From Eq.~(\ref{RA}), firstly, $R_A$ oscillates as a function of $d_{\text{SC}}$ with a period $2\pi/\Delta k$. Secondly, $0\le R_A\le 1/2$, which agrees well with the numerical results shown later. For an illustration, $R_A$ and $R$ are plotted with respect to $d_{\text{SC}}$ for $|\alpha|^2=1-|\beta|^2=0.7$ in Fig.~\ref{Fig1}(d) based on Eq.~(\ref{RA}).

Physically, due to the charge conservation, such a process must have a Cooper pair created and injected into the NSC with a probability $R_A$~\cite{Blonder1982,Entin2008}. The junction therefore has a nonzero conductance when a current $I$ is applied between leads $1$ and $3$ as shown in Fig.~\ref{Fig1}(c). We employ the Landauer-B\"{u}ttiker formula $I_i=(e^2/h)\sum_{j}(T_{ij}V_j-T_{ji}V_i)$ to calculate the conductance, where $I_i$ is the current flowing out of lead $i$, $V_i$ is the voltage of lead $i$, and $T_{ij}$ is the generalized transmission probability from lead $i$ to lead $j$ contributed by both the normal scattering and the Andreev scattering~\cite{Entin2008}. In this $4$-terminal junction, $T_{43}=T_{32}=2R_A=t$ represents the charge transmitted between QAH and NSC, $T_{42}=1-2R_A=r$ is the charge reflected from lead $4$ to $2$~\cite{transmission}, $T_{14}=T_{21}=1$, and all the other $T_{ij}$ are zero. One finds
\begin{equation}
\sigma_{13}=\frac{I}{V_1-V_3}=2R_A\frac{e^2}{h}\ ,\quad \sigma_{24}=\frac{I}{V_2-V_4}=\frac{e^2}{h}\ .
\end{equation}
Therefore, the two-terminal longitudinal conductance $\sigma_{13}$ exhibits an Andreev oscillation with respect to $\phi$, while the Hall conductance $\sigma_{24}$ remains quantized.

\begin{figure}
\begin{center}
\includegraphics[width=3.3in]{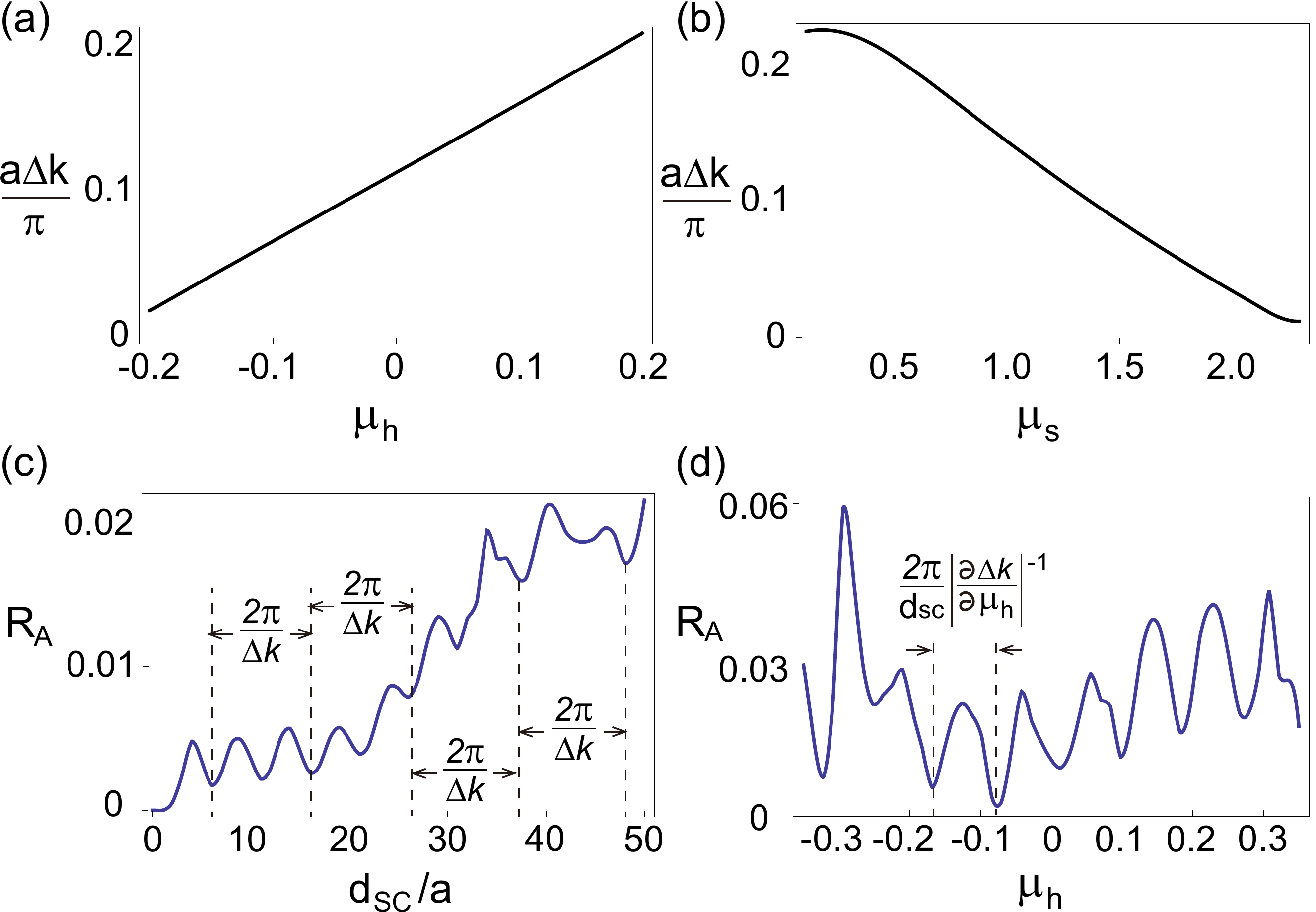}
\end{center}
\caption{(color online). (a) $\Delta k$ as a function of $\mu_h$ with $\mu_s=0.5$. (b) $\Delta k$ as a function of $\mu_s$ with $\mu_h=0.2$. (c) $R_A$ of an edge electron wave packet with respect to $d_{\text{SC}}$ calculated for $(\mu_h,\mu_s)=(0.2,0.8)$. (d) $R_A$ of an edge electron wave packet with respect to $\mu_h$ calculated for $\mu_s=0.8$ and $d_{\text{SC}}=50a$.}
\label{Fig2}
\end{figure}

\begin{figure}
\begin{center}
\includegraphics[width=3.3in]{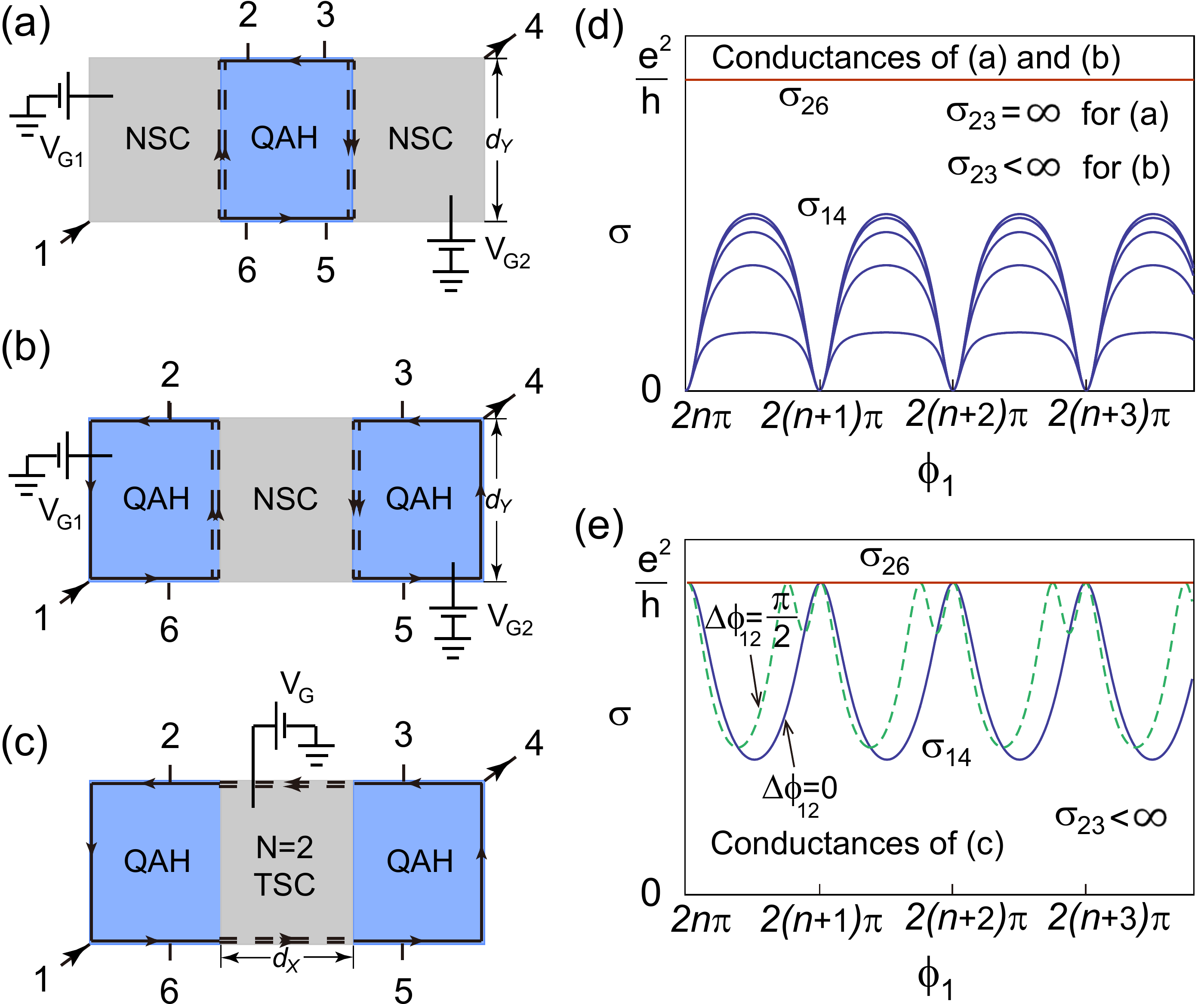}
\end{center}
\caption{(color online). (a)-(c) Illustration of three examples of 6-terminal QAH/SC junctions. (d) $\sigma_{14}$ and $\sigma_{26}$ of junctions (a) and (b) (which are the same) with respect to $\phi_1$ for $\phi_2=n\pi/5(\mbox{mod}\ 2\pi)$ ($1\le n\le5$ from lower to higher). Note their $\sigma_{23}$ are different. (e) $\sigma_{14}$ and $\sigma_{26}$ of junction (c) vs. $\phi_1$ for $\Delta\phi_{12}=0,\pi/2$, respectively.}
\label{Fig3}
\end{figure}

In order to observe the oscillatory $\sigma_{13}$, one needs to tune the phase difference $\phi$. One way is to continuously tune the length $d_{\text{SC}}$ of NSC in contact with QAH, which is not quite feasible in experiments. The other way is to tune the momentum difference $\Delta k$, which can be achieved by tuning the chemical potential of either the QAH or the NSC. Since states $\psi_1$ and $\psi_2$ form a PHS pair, their dispersions will shift oppositely in energy (up and down, respectively) as the chemical potential varies, which results in a change of $\Delta k$. To verify this argument, we have calculated $\Delta k$ numerically as a function of $\mu_h$ and $\mu_s$ for the model and parameters mentioned above, which are presented in Fig.~\ref{Fig2}(a) ($\mu_s=0.5$) and Fig.~\ref{Fig2}(b) ($\mu_h=0.2$), respectively. The results show $\Delta k$ depends almost linearly on $\mu_h$ and $\mu_s$. Thus, one should be able to observe the conductance oscillation by tuning $\mu_h$ or $\mu_s$. As a numerical check, we further calculated the real space evolution of an edge electron wave packet in a low energy window $E\in[-0.1,0.1]$ from lead $4$ to $2$ in the junction, where we chose a lattice size $30\times50$ for the QAH side and $18\times L$ for the NSC side with $0\le L\le50$, and adopted a sine-square deformation to reduce the finite size effect~\cite{supplement,Gendiar2009,Hotta2012}. The contact edge length $d_{\text{SC}}\equiv La$. Fig.~\ref{Fig2}(c) shows $R_A$ as a function of $d_{\text{SC}}$ for $(\mu_h,\mu_s)=(0.2,0.8)$, where one finds the fundamental oscillation period of $2\pi/\Delta k\approx 11a$. We note the $R_A$ oscillation does not reach zero and varies in the amplitude, because $\Delta k$ is dispersive in the energy window of the wave packet. We further plot $R_A$ vs. $\mu_h$ for $\mu_s=0.8$ and $d_{\text{SC}}=50a$ in Fig.~\ref{Fig2}(d), where again one can identify the predicted oscillation period $(2\pi/d_{SC})|\partial\Delta k/\partial \mu_h|^{-1}\approx0.08$. As shown in the supplementary material \cite{supplement}, the oscillation in $R_A$ is robust against disorders. The only difference is that $\Delta k$ will acquire a spatial dependence under disorders, and the phase difference $\phi$ will become $\phi=\int_0^{d_{\text{SC}}}\Delta k\mbox{d}\ell$.

In realistic QAH materials like magnetic (Bi,Sb)$_2$Te$_3$ and graphene, $\Delta k$ usually does not exceed $0.1\pi/a$ with $a$ being the lattice constant. Thus, the spatial oscillation period in $d_{\text{SC}}$ is usually between $10a$ and $10^2a$. The slope $|\partial\Delta k/\partial{\mu_h}|\sim v_F^{-1}\sim0.5$~(eV$\cdot${\AA})$^{-1}$ with $v_F$ the Fermi velocity of the QAH edge state, and $|\partial\Delta k/\partial{\mu_s}|\sim0.1|\partial\Delta k/\partial{\mu_h}|$ is smaller according to our numerical results above. If we take a contact edge length $d_{\text{SC}}=1~\mu$m and tune $\mu_h$ and $\mu_s$, the oscillation periods of $\mu_h$ and $\mu_s$ will be of order of $1$~meV and $10$~meV respectively, in the accessible range of transport experiments. Due to the dispersion of $\Delta k$ in energy, the oscillations become decoherent and invisible above a temperature scale $k_BT\equiv[2\pi /(d_{\text{SC}}|\partial v_F^{-1}/\partial\mu_h|)]^{1/2}$.
Typical values of $|\partial v_F^{-1}/\partial\mu_h|\sim 0.5$~eV$^{-2}$\AA$^{-1}$ and $d_{\text{SC}}=1\mu$m would require $T<300$~K, which is feasible in experiments.

All the above analysis of Majorana fermion interference can be generalized to other QAH/SC junctions. Fig.~\ref{Fig3}(a)-(c) shows three examples of 6-terminal junctions, each of which have two QAH edges proximity-coupled to SC. The chiral Majorana fermions (dashed lines) on these two edges (left and right in junctions (a) and (b), upper and lower in junction (c)) may have distinct phase differences $\phi_1$ and $\phi_2$, and therefore distinct Andreev reflection probabilities $R_{A1}=R_A(\phi_1)$ and $R_{A2}=R_A(\phi_2)$. Junctions (a) and (b) can be implemented by attaching QAH and NSC samples together, while the $N=2$ TSC in junction (c) can be realized via SC proximity on top of the middle region of a QAH sample~\cite{Wang2015}.
With a current $I$ flowing between leads $1$ and $4$, the conductances $\sigma_{ij}=I/(V_i-V_j)$ can be similarly derived from the Landauer-B\"{u}ttiker formula~\cite{supplement}, as listed in Table~\ref{Tab}. The Hall conductance $\sigma_{26}$ is quantized for all the three junctions. In particular, we note that junction (a), which is just the QAH system in a standard Hall bar with SC leads~\cite{lead}, has no difference in $\sigma_{26}$ and $\sigma_{23}$ with the Hall bar with metallic leads. However, $\sigma_{14}$ of such a junction with SC leads is oscillatory with $\phi_1$ and $\phi_2$. In junctions (a) and (b), $\phi_{1}$ and $\phi_2$ can be tuned independently by the gate voltages $V_{G1}$ and $V_{G2}$, respectively. The blue curves in Fig.~\ref{Fig3}(d) show $\sigma_{14}$ vs. $\phi_1$ for fixed $\phi_2=n\pi/5 (\text{mod}\ 2\pi)$ ($1\le n\le5$) and $|\alpha|^2=0.7$. In junction (c), $\phi_{1}$ and $\phi_2$ can be tuned together by the gate voltage $V_{G}$, with $\Delta\phi_{12}\equiv\phi_1-\phi_2$ approximately fixed. In this case, $\sigma_{14}(\phi_1)$ for $\Delta\phi_{12}=0$ and $\pi/2$ are shown in Fig.~\ref{Fig3}(e).

\begin{table}[t]
\renewcommand{\arraystretch}{1.7}
\caption{\label{Tab} The conductances of junctions (a)-(c) shown in Fig.~\ref{Fig3} calculated by the Landauer-B\"{u}ttiker formula.}
\begin{ruledtabular}
\begin{tabular}{cccc}
Junction & $\sigma_{14}$ & $\sigma_{23}$ & $\sigma_{26}$\\
\colrule
(a) & $\frac{2R_{A1}R_{A2}}{R_{A1}+R_{A2}-2R_{A1}R_{A2}}\frac{e^2}{h}$ & $\infty$ & $\frac{e^2}{h}$ \\
(b) & $\frac{2R_{A1}R_{A2}}{R_{A1}+R_{A2}-2R_{A1}R_{A2}}\frac{e^2}{h}$ &$\frac{2R_{A1}R_{A2}}{R_{A1}+R_{A2}-4R_{A1}R_{A2}}\frac{e^2}{h}$ & $\frac{e^2}{h}$\\
(c) & $\frac{R_{A1}+R_{A2}-4R_{A1}R_{A2}}{R_{A1}+R_{A2}-2R_{A1}R_{A2}}\frac{e^2}{h}$ & $\frac{R_{A1}+R_{A2}-4R_{A1}R_{A2}}{2R_{A1}R_{A2}}\frac{e^2}{h}$ & $\frac{e^2}{h}$\\
\end{tabular}
\end{ruledtabular}
\end{table}

\begin{figure}[t]
\begin{center}
\includegraphics[width=3.3in]{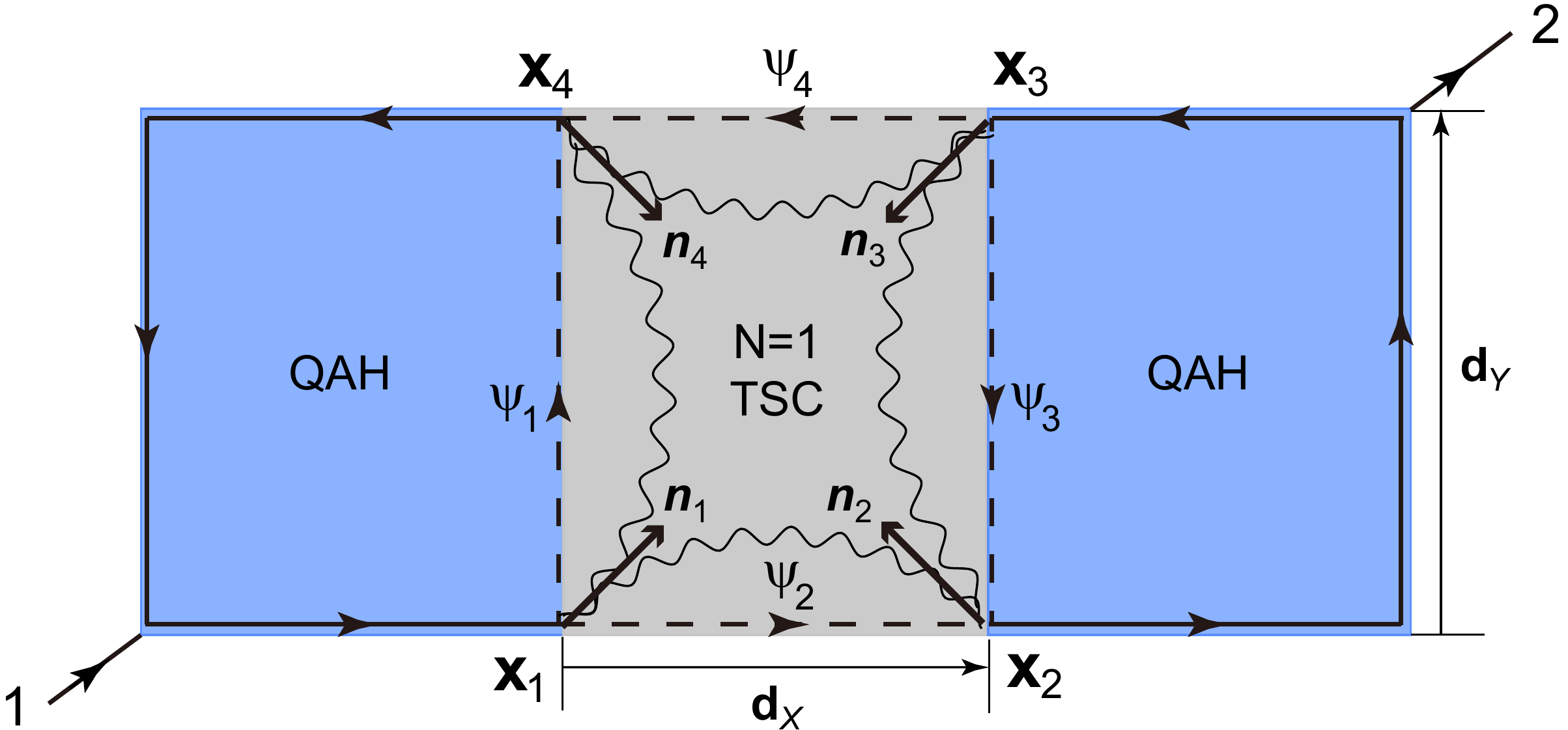}
\end{center}
\caption{Illustration of the QAH/TSC/QAH junction. The fluctuating supercurrent (the wavy lines) contributes a geometry-dependent correction to the conductance $\sigma_{12}$.}
\label{Fig4}
\end{figure}

Finally, we discuss the QAH/TSC/QAH junction as shown in Fig.~\ref{Fig4}, where the TSC has only a single chiral Majorana state $\psi_i$ ($1\le i\le4$) on the $i$-th edge. At the BdG level, an electron incident from lead $1$ will split into $\psi_1$ which is totally reflected and $\psi_2$ which is perfectly transmitted to lead 2, resulting in a half quantized longitudinal conductance $\sigma_{12}=e^2/2h$~\cite{Chung2011,Wang2015}. Here we show when the dynamical fluctuation of the SC phase $\theta$ is considered, $\sigma_{12}$ is no longer exactly quantized but has a geometry-dependent correction $\delta\sigma_{12}$. Such dynamics of the 2D TSC can be described by the effective Hamiltonian~\cite{supplement}
\begin{eqnarray}
&&H_{\text{eff}}=\frac{1}{2g}\int_{\mathcal{M}_{\text{sc}}} \mbox{d}^2\mathbf{x}\left[(\partial_t\theta)^2+v_{s}^2(\nabla\theta)^2\right]
\nonumber
\\
&&-i v_F\sum_{i=1}^{4}\left[\left(\psi_i\psi_{i+1}\mathbf{n}_i\cdot\nabla\theta\right)_{\mathbf{\mathbf{x}}_i}+\int_{\partial_i\mathcal{M}_{\text{sc}}} \mbox{d}\ell \psi_i\partial_{\ell}\psi_i\right] ,
\end{eqnarray}
where $\psi_5\equiv-\psi_1$, $\mathcal{M}_{\text{sc}}$ and $\partial_i\mathcal{M}_{\text{sc}}$ are the bulk and $i$-th edge of the TSC, and the vector potential $\mathbf{A}=0$ gauge is chosen. The Ginzburg-Landau theory gives $g=\mu_0\hbar^2/16m^2\xi^4wB_c^2$ and $v_{s}=\hbar/4m\xi$, where $\mu_0$ is the vacuum permeability, $m$ is the electron effective mass, $\xi$ is the coherence length, $B_c$ is the critical magnetic field~\cite{Bc}, and $w$ is the thickness of the TSC~\cite{Lifshitz1980}. The vector $\mathbf{n}_i$ shown in Fig.~\ref{Fig4} characterizes the interaction between Majorana fermions $\psi_i$ and the supercurrent $\mathbf{j}_s\propto\nabla\theta$ at $\mathbf{x}_i$, and $|\mathbf{n}_i|$ is of the order of the Majorana edge state width. As a result, $\psi_1$ ($\psi_2$) will have a nonzero scattering amplitude into $\psi_3$ ($\psi_4$) via $\mathbf{j}_s$ (wavy lines in Fig.~\ref{Fig4}), leading to a correction to the longitudinal conductance~\cite{supplement}
\begin{equation}
\delta \sigma_{12}\equiv\sigma_{12}-\frac{e^2}{2h}=\frac{e^2}{2h}\frac{g\hbar}{16\pi^2v_s}\sum_{p,q\in\mathbb{Z}}f(p\mathbf{d}_X+q\mathbf{d}_Y),
\end{equation}
where $\mathbf{d}_{X,Y}$ are vectors along the TSC edges as shown in Fig.~\ref{Fig4}. The function $f(\mathbf{x})$ is given by
\begin{equation}
f(\mathbf{x})=\sum_{i,j=1}^4\left(\mathbf{n}_i\cdot\nabla\right)\left(\mathbf{n}_j\cdot\nabla\right) \frac{(-1)^{i-j}(1-\delta_{ij})}{\sqrt{|\mathbf{x}-\mathbf{t}_{ij}|^2+v_s^2|\mathbf{t}_{ij}|^2/v_F^2}},
\nonumber
\end{equation}
where $\mathbf{t}_{ij}$ equals $\mathbf{d}_X/2$ for $i-j$ odd and $\mathbf{d}_Y/2$ for $i-j$ even. Therefore, $\delta\sigma_{12}$ depends on the aspect ratio $\tau=d_Y/d_X$ of the TSC, and scales as $1/d_X^3$ for a fixed $\tau$. In particular, $\delta\sigma_{12}>0$ for $\tau\gg1$, and $\delta\sigma_{12}<0$ for $\tau\ll1$. For a 2D TSC with $w=5$~nm, $\xi=10$~nm, $B_c=0.01$~T and an edge state width $10$~nm, one has $|\delta\sigma_{12}|\sim10^{-6}(e^2/h)$ for $d_{X,Y}\sim1~\mu$m. Therefore, this geometric correction is generically small in experiments.

To conclude, we have proposed transport experiments to detect the Andreev oscillation due to the edge chiral Majorana fermion interference in the QAH/SC junctions. We emphasize that all the conclusions here also apply to ordinary IQH/SC junctions, provided the magnetic field realizing the IQH state is smaller than the upper critical field of the SC. Candidate materials include graphene and Niobium. Moreover, the longitudinal conductance may have multiple oscillation periods if the IQH (QAH) insulator has $N_h>1$ edge chiral fermions, which remains to be studied in details in the future.

\begin{acknowledgements}
We are grateful to Philip Kim for helpful discussions. This work is supported by the US Department of Energy, Office of Basic Energy Sciences, Division of Materials Sciences and Engineering, under Contract No.~DE-AC02-76SF00515 and in part by the NSF under grant No.~DMR-1305677. J.W. is supported by the National Thousand-Young-Talents Program.
\end{acknowledgements}

\bibliography{CMI_ref}

\begin{thebibliography}{39}
\bibitem [{tra()}]{transmission}%
  \BibitemOpen
  \href@noop {} {}\bibinfo {note} {Here we assume a perfect transmission
  between the NSC and the metallic wire attached to lead $3$. If not, one would
  effectively have $t=1-r=2\alpha R_A$ with $0<\alpha<1$.}\BibitemShut {Stop}%
\bibitem [{lea()}]{lead}%
  \BibitemOpen
  \href@noop {} {}\bibinfo {note} {It does not affect the conductances whether
  a voltage lead is superconducting or metallic, since no current flows out of
  the lead.}\BibitemShut {Stop}%
\bibitem [{Bc()}]{Bc}%
  \BibitemOpen
  \href@noop {} {}\bibinfo {note} {For type II SCs
  $B_c\approx\sqrt{B_{c1}B_{c2}}$, where $B_{c1}$ and $B_{c2}$ are the lower
  and upper critical fields.}\BibitemShut {Stop}%
\end{thebibliography}

\begin{widetext}

\section{Supplemental Online Material}

\subsection{Derivation of the Andreev reflection probability $R_A(\phi)$}

In the simplified one-dimensional (1D) picture given by Eq. (3) and the corresponding paragraph of the main text, we have shown the edge chiral wave function $\Psi(\ell)$ of an incident electron is given by
\begin{equation}
\Psi(\ell)=\begin{cases}
\bar{c}(\ell)e^{ik_I\ell}\ , & (\ell<0)\\
\mathcal{N}[\alpha^*\psi_1(\ell)-\beta\psi_2(\ell)]=\mathcal{N}[\left(|\alpha|^2e^{i\Delta k\ell/2}-|\beta|^2e^{-i\Delta k\ell/2}\right)\bar{c}(\ell) +2i\alpha^*\beta\sin(\Delta k\ell/2)\bar{c}^\dag(\ell)]\ , & (0<\ell<d_{SC})\\
u\bar{c}(\ell)e^{ik_I\ell}+v\bar{c}^\dag(\ell)e^{-ik_I\ell}\ , & (\ell>d_{SC})
\end{cases}
\end{equation}
where $\mathcal{N}$ is a normalization factor. However, this 1D wave function $\Psi(\ell)$ cannot be continuous simultaneously at $\ell=0$ and $\ell=d_{SC}$. This is due to the fact that the edge chiral wave function is intrinsically a 2D wave function (which is continuous) and does not exist in 1D systems. To make our 1D picture work, we relax the junction conditions at $\ell=0$ and $\ell=d_{SC}$ as $\Psi(0^+)\propto\Psi(0^-)$ and $\Psi(d_{SC}^+)\propto\Psi(d_{SC}^-)$, where $\ell^\pm$ denotes the right/left limit of position $\ell$. The condition at $\ell=0$ is already satisfied, while that at $\ell=d_{SC}$ requires
\[\frac{ue^{ik_Id_{SC}}}{|\alpha|^2e^{i\phi/2}-|\beta|^2e^{-i\phi/2}}=\frac{ve^{-ik_Id_{SC}}}{2i\alpha^*\beta\sin(\phi/2)}\ .\]
Together with the unitarity condition $|u|^2+|v|^2=1$ we find
\begin{equation}
(u,v)=\frac{\left(e^{-ik_Id_{SC}}(|\alpha|^2e^{i\phi/2}-|\beta|^2e^{-i\phi/2}),\ 2i\alpha^*\beta\sin(\phi/2) e^{ik_Id_{SC}}\right)}{\left[(|\alpha^2|-|\beta^2|)^2+8|\alpha\beta|^2\sin^2(\phi/2)\right]^{1/2}}\ .
\end{equation}
Therefore, we find $R_A(\phi)=|v|^2$ as given in Eq. (4) of the main text.

\begin{figure}[b]
\begin{center}
\includegraphics[width=3.5in]{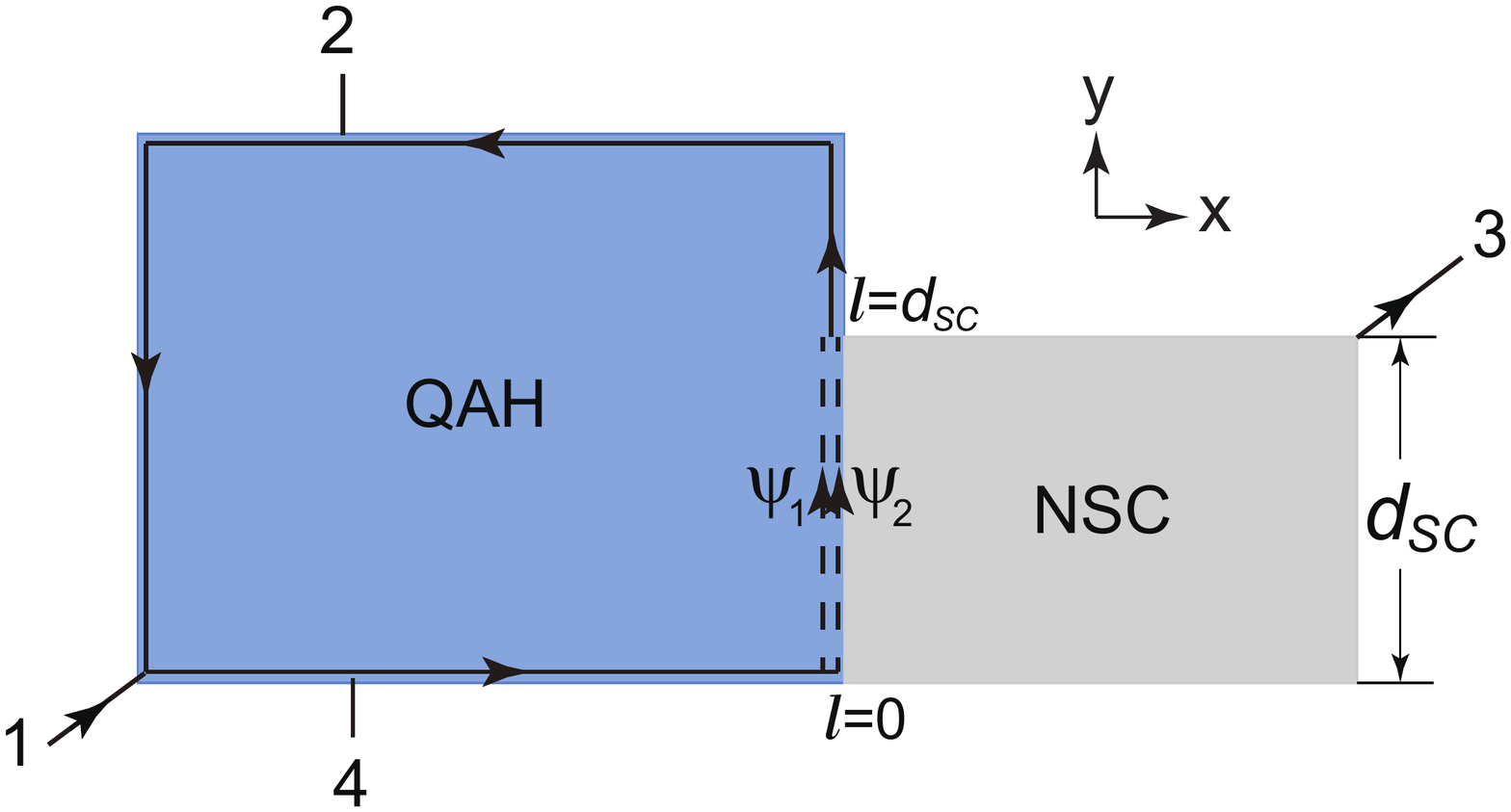}
\end{center}
\caption{(color online). Junction geometry used in the 2D lattice numerical calculations.
}\label{FigS}
\end{figure}

\subsection{2D lattice numerical calculations for the QAH/NSC junction}

To verify the $R_A$ oscillation induced by the momentum difference $\Delta k$, we calculate the propagation of an edge electron wave packet in a 2D lattice QAH/NSC junction as shown in Fig. \ref{FigS}, and use the model and parameters as presented at the beginning of the main text. The size of the QAH lattice is $30\times50$, while that of the NSC lattice is $18\times L$ with $0\le L\le 50$. We therefore have $d_{SC}=La$. The wave packet is restricted inside an in-gap low energy window $E\in[-0.1,0.1]$, and is initially localized at the lower QAH edge around lead 4. After a certain time $t$, the wave packet will propagate to the upper QAH edge around lead 2 and become a superposition of electron state and hole state. We then extract the hole probability as the Andreev reflection probability $R_A$.

To reduce the finite size effect and prevent the wave packet from flowing back via the left QAH edge, we employ the sine-square deformation technique \cite{Gendiar2009,Hotta2012}, which is to deform the Hamiltonian $H(x,y)$ at position $(x,y)$ into $H(x,y)\sin^2(\pi x/L_x)$ where $L_x$ is the system size in the $x$ direction. This makes the hopping on the left QAH edge (and the right edge of NSC) zero, so that the wave packet cannot propagate back to lead 4 from lead 2. Physically, this method simulates the effect of the conducting wires at lead 1 and lead 3.

The oscillation of $R_A$ with respect to $L$ for $(\mu_h,\mu_s)=(0.2,0.8)$ is shown in Fig. 2(c) of the main text. For fixed $L=50$ and fixed $\mu_s=0.8$, the oscillation of $R_A$ with respect to $\mu_h$ is as shown in Fig. 2(d) of the main text. This oscillation is robust against disorders, as the chiral Majorana edge states inducing the oscillation are topologically protected. To see this, we have done another numerical calculation in Fig. \ref{FigS2}(a) below with disorders in chemical potentials $\mu_h,\mu_s$ and pairing amplitude $\Delta_s$ added. The contact length is fixed at $L=50$, the average SC pairing amplitude is $\bar{\Delta}_s=0.3$, and the average SC chemical potential is $\bar{\mu}_s=0.8$. The chemical potential on each site $i$ is $\mu_h(i)=\bar{\mu}_h+\delta\mu_i$ for the QAH side and $\mu_s(i)=\bar{\mu}_s+\delta\mu_i$ for the NSC side, where $\delta \mu_i$ is a random potential obeying the Gaussian distribution with standard deviation $\sigma_\mu=0.05\sim$ $1/10$ of the QAH gap. The pairing amplitude on the NSC side is $\Delta_s(i)=\bar{\Delta}_s+\delta\Delta_i$, where $\delta\Delta_i$ obeys the Gaussian distribution with standard deviation $\sigma_\Delta=0.1\Delta_s=0.03$. The chemical potential disorder $\delta\mu_i$ is fixed throughout the calculation to simulate static inhomogeneities. On the other hand, the pairing fluctuation $\delta\Delta_i$ is regenerated before the calculation of $R_A$ at each $\bar{\mu}_h$, and for each $\bar{\mu}_h$ we calculate $R_A$ three times and take the average, so that the results simulate a dynamical fluctuating pairing amplitude $\Delta_s$. In comparison with the homogeneous result shown in Fig. 2(d) of the main text, one sees that the $R_A$ oscillation pattern is quite robust under disorders.

Such $R_A$ oscillation can also be seen by tuning the SC chemical potential $\mu_s$. Fig. \ref{FigS2}(b) in the below shows $R_A$ vs. $\mu_s$ with fixed $L=50$ and $\mu_h=0.25$. Though the amplitude varies a lot with respect to $\mu_s$, we can see an oscillation pattern in agreement with the predicted oscillation period $(2\pi/d_{SC})|\partial\Delta k/\partial \mu_s|^{-1}$. The oscillation in both $\mu_h$ and $\mu_s$ will become clearer if the system size $d_{SC}$ is larger, because more oscillation periods will be seen, as will be the case in the experiments.

\begin{figure}[b]
\begin{center}
\includegraphics[width=5in]{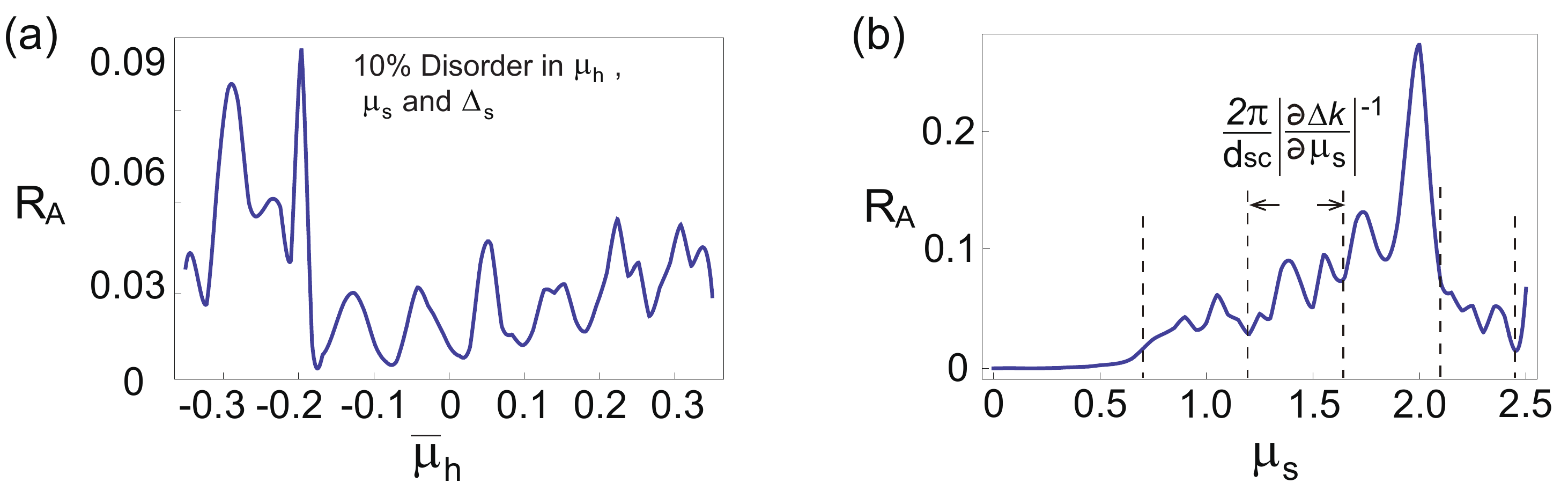}
\end{center}
\caption{(color online). (a) $R_A$ vs. $\bar{\mu}_h$ calculated with static disorders of $10\%$ QAH gap in chemical potentials $(\mu_h,\mu_s)$ and $10\%$ dynamical fluctuations of the SC pairing amplitude $\Delta_s$, at fixed $d_{SC}=50a$ and $\bar{\mu}_s=0.8$. The oscillation pattern is topologically robust when compared to the homogeneous result shown in Fig. 2(d) of the main text. (b) $R_A$ of an edge electron wave packet with respect to $\mu_s$ calculated for homogeneous crystals with $\mu_h=0.25$ and $d_{\text{SC}}=50a$.
}\label{FigS2}
\end{figure}

\subsection{Derivation of the conductivities with the Landauer-B\"{u}ttiker formula}

The conductances of the 6-terminal junction shown in Fig. 3(a) can be easily calculated by writing down the transmission coefficients which we denote as $T_{ij}^{(a)}$:
\begin{equation}
T_{16}^{(a)}=T_{21}^{(a)}=2R_{A1}\ ,\ T_{26}^{(a)}=1-2R_{A1}\ ,\ T_{43}^{(a)}=T_{54}^{(a)}=2R_{A2}\ ,\ T_{53}^{(a)}=1-2R_{A2}\ ,\ T_{65}^{(a)}=T_{32}^{(a)}=1\ ,
\end{equation}
and all the other $T_{ij}^{(a)}=0$. The current is given by $I_4=-I_1=I$ and $I_{2,3,5,6}=0$. The Landauer-B\"{u}ttiker formula $I_i=(e^2/h)\sum_{j}(T_{ij}V_j-T_{ji}V_i)$ then yields
\[
\begin{split}
&I=\frac{e^2}{h}2R_{A_1}(V_1-V_6)=\frac{e^2}{h}2R_{A_2}(V_3-V_4)\ ,\\
&0=2R_{A_1}V_1+(1-2R_{A_1})V_6-V_2=2R_{A_2}V_4+(1-2R_{A_2})V_3-V_5=V_5-V_6=V_2-V_3\ .
\end{split}
\]
If we set $V_1=0$, we find
\[V_5=V_6=-\frac{1}{2R_{A1}}\frac{h}{e^2}I\ ,\ V_2=V_3=-\frac{1-2R_{A1}}{2R_{A1}}\frac{h}{e^2}I\ ,\ V_4=-\frac{R_{A1}+R_{A2}-2R_{A1}R_{A2}}{2R_{A1}R_{A2}}\frac{h}{e^2}I\ .\]
Therefore, the conductances of junction (a) is
\begin{equation}
\sigma_{14}=\frac{I}{V_1-V_4}=\frac{2R_{A1}R_{A2}}{R_{A1}+R_{A2}-2R_{A1}R_{A2}}\frac{e^2}{h}\ ,\ \sigma_{23}=\frac{I}{V_2-V_3}=\infty\ ,\ \sigma_{26}=\frac{I}{V_2-V_6}=\frac{e^2}{h}\ .
\end{equation}

The transmission coefficients $T_{ij}^{(b)}$ of junction (b) are not so straightforward. A cooper pair in the NSC has a probability $r_1=1-2R_{A1}$ ($r_2=1-2R_{A2}$) to be reflected by the left (right) edge. Accordingly, the transmission probability at the left (right) edge is $t_1=2R_{A1}$ ($t_2=2R_{A2}$). Therefore, we have
\[T_{65}^{(b)}=T_{32}^{(b)}=t_1\left[\sum_{n=0}^{\infty}(r_2r_1)^n\right]t_2=\frac{t_1t_2}{1-r_1r_2} =\frac{2R_{A1}R_{A2}}{R_{A1}+R_{A2}-2R_{A1}R_{A2}}\ .\]
Similarly, one finds
\[T_{62}^{(b)}=T_{35}^{(b)}=1-T_{65}^{(b)}=\frac{R_{A1}+R_{A2}-4R_{A1}R_{A2}}{R_{A1}+R_{A2}-2R_{A1}R_{A2}}\ ,\ \ \ \  T_{16}^{(b)}=T_{54}^{(b)}=T_{43}^{(b)}=T_{21}^{(b)}=1\ ,\]
and all the other coefficients are zero. By setting $V_1=0$ and solving the equations, we find
\[V_2=V_1=0\ ,\ V_6=-\frac{h}{e^2}I\ ,\ V_5=V_4=-\frac{R_{A1}+R_{A2}-2R_{A1}R_{A2}}{2R_{A1}R_{A2}}\frac{h}{e^2}I\ ,\ V_3=-\frac{R_{A1}+R_{A2}-4R_{A1}R_{A2}}{2R_{A1}R_{A2}}\frac{h}{e^2}I\ .\]
So the conductances of junction (b) are given by
\begin{equation}
\sigma_{14}=\frac{2R_{A1}R_{A2}}{R_{A1}+R_{A2}-2R_{A1}R_{A2}}\frac{e^2}{h}\ ,\qquad \sigma_{23}=\frac{2R_{A1}R_{A2}}{R_{A1}+R_{A2}-4R_{A1}R_{A2}}\frac{e^2}{h}\ ,\qquad \sigma_{26}=\frac{e^2}{h}\ .
\end{equation}

Junction (c) is quite analogous to junction (b), except that the transmission coefficients become
\[T_{62}^{(c)}=T_{35}^{(c)}=1-T_{65}^{(c)}=1-T_{32}^{(c)}=t_1\left[\sum_{n=0}^{\infty}(r_2r_1)^n\right]t_2=\frac{2R_{A1}R_{A2}}{R_{A1}+R_{A2}-2R_{A1}R_{A2}}\ .\]
As a result, the conductances of junction (c) are
\begin{equation}
\sigma_{14}=\frac{R_{A1}+R_{A2}-4R_{A1}R_{A2}}{R_{A1}+R_{A2}-2R_{A1}R_{A2}}\frac{e^2}{h}\ ,\qquad \sigma_{23}=\frac{R_{A1}+R_{A2}-4R_{A1}R_{A2}}{2R_{A1}R_{A2}}\frac{e^2}{h}\ ,\qquad \sigma_{26}=\frac{e^2}{h}\ .
\end{equation}

\subsection{Contribution of the dynamical SC phase fluctuation to the conductance of a QAH/TSC/QAH junction}

It can be shown that $H_{eff}$ in Eq. (6) of the main text is the only gauge invariant Hamiltonian one can write down for the QAH/TSC/QAH junction of Fig. 4 in the main text. We have defined $\psi_5=-\psi_1$ when writing the interactions since fermions are known to satisfy the anti-boundary condition on a 1D edge when the enclosed flux is zero.
The interaction $H_I=v_F\sum_i(\psi_i\psi_{i+1}\mathbf{n}_i\cdot\nabla\theta)_{\mathbf{x}_i}$ corresponds to the process in which a normal current $\mathbf{j}_n\propto\psi_i\psi_{i+1}\mathbf{n}_i$ on the edge turns into a supercurrent $\mathbf{j}_s\propto\nabla\theta$ in the bulk TSC. Microscopically, the vector coupling strength $\mathbf{n}_i$ which has a dimension of length is given by
\begin{equation}
\mathbf{n_i}=\frac{i}{2v_F}\int_{\mathcal{M}_{SC}} \mbox{d}^2\mathbf{x}\varphi_{i}(\mathbf{x})^\dag\hat{\mathbf{j}}(\mathbf{x})\varphi_{i+1}(\mathbf{x})\ ,
\end{equation}
where $\varphi_{i}(\mathbf{x})$ is the 2D wave function of the edge chiral Majorana mode $\psi_i$ at zero energy (which is a plane wave in the edge direction), and \[\hat{\mathbf{j}}(\mathbf{x})=i\frac{\delta H_{TSC}}{\delta \nabla c(\mathbf{x})}c(\mathbf{x})-i\frac{\delta H_{TSC}}{\delta \nabla c^\dag(\mathbf{x})}c^\dag(\mathbf{x})\]
is the fermion current operator, with $H_{TSC}$ the BdG Hamiltonian of the TSC. The integration mainly comes from the vicinity of $\mathbf{x}_i$ where the two Majorana wave functions overlap (within a radius of the edge state width $\xi$). As a result, $\mathbf{n}_i$ points more or less along the bisector of the angle formed by the two edges, and its norm $|\mathbf{n}_i|$ is of order of the edge state width $\xi$.

When a current is flowing from lead 1 to lead 2, it will enter the TSC at $\mathbf{x}_{1}$ or $\mathbf{x}_{4}$, and leave the TSC at $\mathbf{x}_{2}$ or $\mathbf{x}_{3}$. To determine the conductivity of the junction, we need to calculate the scattering matrix between the charged edge states at the four corners $\mathbf{x}_i$ of the TSC. According to the edge state chirality given in Fig. 4 of the main text, the basis of the incident edge states is $\Psi_{in}=(\bar{c}_1,\bar{c}_3,\bar{c}^\dag_1,\bar{c}^\dag_3)^T$, and the basis of the outgoing edge states is $\Psi_{out}=(\bar{c}_2,\bar{c}_4,\bar{c}^\dag_2,\bar{c}^\dag_4)^T$, where $\bar{c}_i$ annihilates the edge chiral electron on the QAH/vacuum edge that is connected to the corner $\mathbf{x}_i$ of the TSC. They are related to the four Majorana edge states $\psi_i$ in the following way:
\begin{equation}\label{charge}
\bar{c}_1=\frac{\psi_1+i\psi_2}{\sqrt{2}}\ ,\quad \bar{c}_2=\frac{\psi_3+i\psi_2}{\sqrt{2}}\ ,\quad \bar{c}_3=\frac{\psi_3+i\psi_4}{\sqrt{2}}\ ,\quad \bar{c}_4=\frac{\psi_1-i\psi_4}{\sqrt{2}}\ .
\end{equation}
Therefore, to find out the scattering matrix, we need to calculate the scattering amplitude $\mathcal{M}_{ij}$ between Majorana states $\psi_j$ and $\psi_i$ given by
\[2\pi\delta(k-k')\mathcal{M}_{ij}(k)=\lim_{t\rightarrow\infty}\langle \psi_{i,k'}e^{-iH_{eff}t}\psi_{j,-k}\rangle=\frac{\langle \psi_{i,k'}\ T\text{exp}\Big[-i\int_\infty^\infty H_I(t)dt\Big]\psi_{j,-k}\rangle}{\langle T\text{exp}\Big[-i\int_\infty^\infty H_I(t)dt\Big]\rangle}\ ,\]
where $k>0$ and $k'>0$ are the incident and outgoing momentum of the edge Majorana state, $\psi_{i,k}=\int \mbox{d}x\psi_i(x)e^{-ikx}$, $\{\psi_{i,k},\psi_{j,k'}\}=2\pi\delta_{ij}\delta(k+k')$, and $T$ stands for time ordering. The particle vacuum is given by $\psi_{i,k}|0\rangle=0$ for all $k>0$. It is easy to see that to the lowest order $\mathcal{M}_{ij}(k)=\delta_{ij}$. If we keep up to the second order, $\mathcal{M}_{13}$ and $\mathcal{M}_{24}$ will become nonzero. The first part of $\mathcal{M}_{13}$ comes from the scattering from $\psi_1$ to $\psi_3$ via $\psi_2$, which is given by
\[
2\pi\delta(k-k')\mathcal{M}_{13}^{(1)}(k)=v_F^2\langle \psi_{3,k'} \int dt' \int dt \Big[ T \psi_2(\mathbf{x}_2,t')\psi_3(\mathbf{x}_2,t')\mathbf{n}_2\cdot \nabla\theta(\mathbf{x}_2,t') \psi_1(\mathbf{x}_1,t)\psi_2(\mathbf{x}_1,t)\mathbf{n}_1\cdot \nabla\theta(\mathbf{x}_1,t) \Big]  \psi_{1,-k}\rangle\ .
\]
We shall calculate this part first as an illustration. By defining the Green's functions for the Majorana fermion and the supercurrent
\[G_j(\mathbf{x},\mathbf{x}',t)=-i\langle T\psi_j(\mathbf{x},0)\psi_j(\mathbf{x}',t) \rangle\ , \qquad D(\mathbf{x},\mathbf{x}',t)=-i\langle T\theta(\mathbf{x},0)\theta(\mathbf{x}',t)\rangle\ ,\]
one finds
\begin{equation}\label{M13}
\mathcal{M}_{13}^{(1)}(k)=-e^{-ikd_X}v_F\int\frac{d\omega}{2\pi} G_2(\mathbf{x}_2,\mathbf{x}_1,\omega)D_{21}(\mathbf{x}_2,\mathbf{x}_1,-\omega-v_Fk)\ ,
\end{equation}
where the integration is in the frequency $\omega$ space, and we have defined $D_{ij}(\mathbf{x},\mathbf{x}',t)=(\mathbf{n}_i\cdot\nabla_{\mathbf{x}})(\mathbf{n}_j\cdot\nabla_{\mathbf{x}'})D(\mathbf{x},\mathbf{x}',t)$. The Green's function $G_2$ can be readily calculated:
\[G_2(\mathbf{x}_2,\mathbf{x}_1,\omega)=-\int\frac{dk}{2\pi}\frac{e^{ik(x_1-x_2)}}{\omega-v_Fk+i\delta \text{sgn}(k)}=i\frac{e^{-i\omega(x_1-x_2)/v_F}}{v_F}[\Theta(x_1-x_2)\Theta(\omega)-\Theta(x_2-x_1)\Theta(-\omega)]\ ,\]
where $\Theta(x)$ is the Heavyside's step function. In our setup here, $x_2-x_1=d_X>0$. The Green's function $D$ can be easily calculated if the TSC is an infinite 2D plane without a boundary, which we shall call $D^{(0)}$:
\[D^{(0)}(\mathbf{x}',\mathbf{x},\omega) =\int\frac{d^2\mathbf{k}}{(2\pi)^2}\frac{ge^{i\mathbf{k}\cdot(\mathbf{x}-\mathbf{x}')}}{\omega^2-\omega_k^2+i\delta}=-\frac{g}{2\pi v_s^2}J_0(-i\omega |\mathbf{x}'-\mathbf{x}|/v_s)\ ,\]
where $J_0(x)$ is the zeroth Bessel function. For a rectangular TSC bounded by four edges in our setup, the Green's function $D$ can be calculated by the method of images as
\[D(\mathbf{x}',\mathbf{x},\omega)=\sum_{\mathbf{x}^{(I)}}D^{(0)}(\mathbf{x}',\mathbf{x}^{(I)},\omega)\ ,\]
where $\mathbf{x}^{(I)}$ runs over the infinite images of point $\mathbf{x}$ including itself.

Now we can proceed to calculate $\mathcal{M}_{13}^{(1)}(k)$. Since we are interested in the low energy scattering, we shall take the limit $k\rightarrow0^+$ in Eq. (\ref{M13}) and denote $\mathcal{M}_{13}^{(1)}(k\rightarrow0^+)=\mathcal{M}_{13}^{(1)}$. Therefore, we find
\begin{equation}
\mathcal{M}_{13}^{(1)}=-v_F\sum_{\mathbf{x}_1^{(I)}}(\mathbf{n}_1\cdot\nabla_{\mathbf{x}_1})(\mathbf{n}_2\cdot\nabla_{\mathbf{x}_2}) \int\frac{d\omega}{2\pi} G_2(\mathbf{x}_2,\mathbf{x}_1,\omega)D^{(0)}(\mathbf{x}_2,\mathbf{x}_1^{(I)},-\omega)
=\frac{g\hbar}{16\pi^2v_s}\sum_{p,q\in\mathbb{Z}}f_{12}(p\mathbf{d}_X+q\mathbf{d}_Y)\ ,
\end{equation}
where
\[f_{12}(\mathbf{x})=-\left(\mathbf{n}_1\cdot\nabla\right)\left(\mathbf{n}_2\cdot\nabla\right) \frac{2}{\sqrt{|\mathbf{x}-(\mathbf{d}_X/2)|^2+v_s^2|\mathbf{d}_X|^2/4v_F^2}}\ .\]
Similarly, one can calculate the second part $\mathcal{M}_{13}^{(2)}$ coming from the scattering from $\psi_1$ to $\psi_3$ via $\psi_4$. The total scattering amplitude is then $\mathcal{M}_{13}=\mathcal{M}_{13}^{(1)}+\mathcal{M}_{13}^{(2)}$. The calculation of $\mathcal{M}_{24}$ follows the same procedure.

When rewritten in terms of the charge basis in Eq. (\ref{charge}), we find the scattering matrix to be
\begin{equation}
\left(\begin{array}{c}\bar{c}_2\\ \bar{c}_4\\ \bar{c}_2^\dag\\ \bar{c}_4^\dag\end{array}\right)=\frac{1}{2}
\left(\begin{array}{cccc}
1+\mathcal{M}_{13}&1-\mathcal{M}_{24}&-1+\mathcal{M}_{13}&1+\mathcal{M}_{24}\\ 1-\mathcal{M}_{24}&-1-\mathcal{M}_{13}&1+\mathcal{M}_{24}&1-\mathcal{M}_{13}\\ -1+\mathcal{M}_{13}&1+\mathcal{M}_{24}&1-\mathcal{M}_{13}&1-\mathcal{M}_{24}\\ 1+\mathcal{M}_{24}&1-\mathcal{M}_{13}&1-\mathcal{M}_{24}&-1-\mathcal{M}_{13} \end{array}\right)
\left(\begin{array}{c}\bar{c}_1\\ \bar{c}_3\\ \bar{c}_1^\dag\\ \bar{c}_3^\dag\end{array}\right)\ ,
\end{equation}
where we have assumed $\mathcal{M}_{13}$ and $\mathcal{M}_{24}$ are small. Accordingly, we find the normal transmission probability ($\bar{c}_1\rightarrow\bar{c}_2$) to be $\mathcal{T}=|1+\mathcal{M}_{13}|^2/4$, and the Andreev reflection probability (($\bar{c}_1\rightarrow\bar{c}_4^\dag$)) to be $\mathcal{R}_A=|1+\mathcal{M}_{24}|^2/4$. Therefore, we find the longitudinal conductance \cite{Chung2011,Wang2015}
\[\sigma_{12}=\frac{e^2}{h}(\mathcal{T}+\mathcal{R}_A)\approx\frac{e^2}{2h}(1+\mathcal{M}_{13}+\mathcal{M}_{24})\ .\]
The correction is given by
\begin{equation}
\delta \sigma_{12}=\frac{e^2}{2h}(\mathcal{M}_{13}+\mathcal{M}_{24}) =\frac{e^2}{2h}\frac{g\hbar}{16\pi^2v_s}\sum_{p,q\in\mathbb{Z}}f(p\mathbf{d}_X+q\mathbf{d}_Y) \ ,
\end{equation}
where
\[f(\mathbf{x})=\sum_{i,j=1}^4\left(\mathbf{n}_i\cdot\nabla\right)\left(\mathbf{n}_j\cdot\nabla\right) \frac{(-1)^{i-j}(1-\delta_{ij})}{\sqrt{|\mathbf{x}-\mathbf{t}_{ij}|^2+v_s^2|\mathbf{t}_{ij}|^2/v_F^2}}\ .\]

\end{widetext}

\end{document}